\documentclass[prb,aps,showpacs,prb,amsmath,floatfix,showkeys,twocolumn]{revtex4}
\usepackage[]{graphicx}
\usepackage{graphicx}
\usepackage{dcolumn}
\usepackage{bm}

\begin{document}
\title{Real time electron dynamics in an interacting
vibronic molecular quantum dot}
\author{A. Goker$^{1}$}

\affiliation{$^1$
Department of Physics, Bilecik University,  \\
Gulumbe, 11210, Bilecik, Turkey
}

\date{\today}

\begin{abstract}
We employ the time-dependent non-crossing approximation 
to investigate the joint effect of strong electron-electron 
and electron-phonon interaction on the instantaneous 
conductance of a single molecule transistor which is 
abruply moved into the Kondo regime by means of a gate 
voltage. We find that the instantaneous conductance 
exhibits decaying sinusoidal oscillations in the long 
timescale for infinitesimal bias. Ambient temperature
and electron-phonon coupling strength influence the
amplitude of these oscillations. The frequency of 
oscillations is found to be equal to the phonon frequency. 
We argue that the origin of these oscillations can 
be attributed to the interference between the emerging 
Kondo resonance and its phonon sidebands. We discuss the 
effect of finite bias on these oscillations.    
\end{abstract}



\thispagestyle{headings}

\maketitle

\section{Introduction}

Molecular transport junctions that consist 
of a molecule inserted between contacts long 
fascinated both physicists and chemists since 
it has been proposed more than three decades 
ago that they could be used one day as building 
blocks for future electronic devices. The 
emergence of the field of molecular electronics 
originates from this suggestion \cite{AvirametAl74CPL}. 
The nanotechnology revolution which took place 
in mid 1980's paved the way for precise control 
on nanostructures and enabled to perform the 
experiments that were beyond anyone's 
realm previously \cite{KrotoetAl85Nature}. 
Experimental confirmation of the Kondo effect, 
which was predicted two decades ago theoretically 
\cite{NgLee88PRL,GlazmanRaikh88JL}, first on quantum dots  
\cite{GoldhaberetAl98Nature,GoldhaberetAl98PRL,CronenwettetAl98Science} 
and then on single molecule transistors 
\cite{ParketAl02Nature,LiangetAl02Nature} opened 
an avenue for the possible merger of the fields of 
spintronics and molecular electronics \cite{RochaetAl05NatureMat}.
The goal of the field of molecular spintronics is 
to investigate spin-dependent transport in 
molecular electronic devices \cite{Seneoretal07JPCM}.
Therefore, the Kondo effect acts as a supplemental
spin-dependent mechanism that might strongly affect
the spin transport in quantum dots.    
 
In his groundbreaking work \cite{Kondo64PTP}, 
Jun Kondo discovered that resistivities of 
bulk metals which contain magnetic impurities 
with localized unpaired spins would be enhanced 
at low temperatures. This enhancement was dubbed 
as the Kondo effect later on. The underlying 
microscopic mechanism for the Kondo effect has 
been identified as the formation of a spin singlet 
resulting from the interaction of the unpaired 
localized electron and continuum electrons 
near the Fermi level in the metal \cite{Hewson93BOOK}.

Quantum dots are artificial atoms that can 
contain an integer number of electrons 
\cite{Ashoori96Nature}. It is imperative to 
confine odd number of electrons in a quantum 
dot to observe the Kondo effect. This gives 
rise to a net spin in the dot just like in a 
bulk metal. However, coupling of this net spin 
to the fermionic bath in the metallic leads 
opens a new transport channel in the quantum 
dot and conductance is enhanced at low 
temperatures instead of resistivity in bulk metals.
A sharp many-body resonance at the dot density of 
states pinned to the Fermi level of the contacts 
is responsible for the enhancement. This 
conductance enhancement is a robust way of 
obtaining current flow in odd number Coulomb 
blockade valleys and thus to prevent current 
inhibition in single electron devices.

Sudden shifting of the gate or source-drain 
voltage has been studied in detail previously
\cite{NordlanderetAl99PRL,PlihaletAl00PRB,
SchillerHershfield00PRB,MerinoMarston04PRB} and 
three unique timescales have been clearly 
identified in the ensuing transient current 
\cite{PlihaletAl05PRB,AndersSchiller05PRL,AndersetAl06PRB,
IzmaylovetAl06JPCM}. Non-Kondo timescale is associated with 
the initial fast rise of the current accompanied by 
reshaping of the Breit-Wigner resonance. On the other hand, 
formation of the Kondo resonance and its reaching a broad
quasismooth structure with a linewidth on the order of 
Kondo temperature is the hallmark of longer Kondo 
timescale. Splitting of the Kondo resonance for finite bias 
corresponds to the third and longest timescale.
This timescale can be determined by measuring the 
decay rate of the split Kondo peak oscillations 
\cite{PlihaletAl05PRB}. Later studies focused on the 
asymmetric coupling of the dot to the contacts and
concluded that an interference between the Kondo resonance 
and the van Hove singularities in the density of states of 
the leads may give rise to sinusoidal oscillations in the 
transient current \cite{GokeretAl07JPCM}. Recent extension 
of the diagrammatic Monte Carlo method \cite{GulletAl08EPL}
to nonequilibrium impurity problems \cite{WerneretAl09PRB}
confirmed the sensitive dependency of the transient current
on the bandwidth of the contacts \cite{SchmidtetAl08PRB}.

The history of electron-phonon interaction in molecular
transport junctions is quite long as it has been 
summarized in a recent review \cite{GalperinetAl07JPCM}. 
Phonon assisted electron transport in molecular transport 
junctions can be broadly classified based on the relative 
time and energy scales in the process. The strength of the 
electron-phonon interaction is judged relative to the 
molecule-electrode coupling. In this respect, weak 
and strong electron-phonon coupling regimes emerge.

The former regime, namely the weak electron-phonon 
interaction, corresponds to nonresonant phonon assisted 
electron tunneling encountered in inelastic electron 
transport spectroscopy \cite{StipeetAl99PRL,HahnetAl00PRL}. 
The development of scanning tunneling microscope and 
spectroscopy proved to be an invaluable tool for inelastic 
electron transport spectroscopy to define and characterize 
the conductance properties of molecular species. 
It is justified to use Migdal-Eliashberg theory 
\cite{Migdal58JETP,Eliashberg60JETP} in this regime. As 
a result, second order perturbation theory on electron-phonon 
coupling over the Keldysh contour leads to Born approximation. 
Various versions of the self-consistent Born approximation 
has been used in several theoretical studies
\cite{FrederiksenetAl04PRL,MiietAl03PRB,GalperinetAl04NL,GalperinetAl04JCP}.

The latter regime corresponds to resonant tunneling 
which involves longer electron lifetime and stronger 
electron-phonon interaction. Perturbation theory fails 
in this regime and polaron is formed in the junction. 
Signatures of the resonant tunneling are manifested as 
sidebands of the main Kondo resonance in differential 
conductance \cite{YuetAl04NL}. There are several 
studies investigating this case in steady state. 
Preliminary accounts of the electron-phonon coupling 
computed the dot Green's function using the equation of 
motion method \cite{ZhuetAl03PRB} and treating 
the contacts as unaffected by the bosons in the wide band 
limit \cite{LundinetAl02PRB}. They both concluded 
that the electron-phonon interaction results in sidebands only 
in one side of the main elastic peak in density of states. 
In later studies, it has been suggested that this is due to 
an invalid approximation employed in calculating the retarded 
Green's function and the phonon sidebands should appear in 
both sides of the elastic peak \cite{ChenetAl05PRB,GalperinetAl06PRB,WangetAl07PRB}. 
Perturbative renormalization group calculations confirmed 
this latter conclusion \cite{PaaskeetAl05PRL}.

In strong electron-phonon coupling coupling regime, the 
approach invoking Lang-Firsov canonical transformation and 
non-crossing approximation also found sidebands in both sides 
of the main Kondo resonance in time averaged ac conductance, 
however this last method concluded that the zero bias Kondo 
resonance has been greatly suppressed \cite{YongetAl07CTP}. 
Recently, investigation of the electron-phonon interaction 
in Kondo regime using nonequilibrium equation of motion 
method has found that increasing electron-phonon interaction 
strength gradually destroys the Kondo effect \cite{GalperinetAl07PRB}. 
This has been attributed to the destruction of the coherence 
in the system by the electron-phonon interaction and the shift 
of the energy level due to rearrangement of the phonons.

\begin{figure}[htb]
\centerline{\includegraphics[angle=0,width=8.0cm,height=5.8cm]{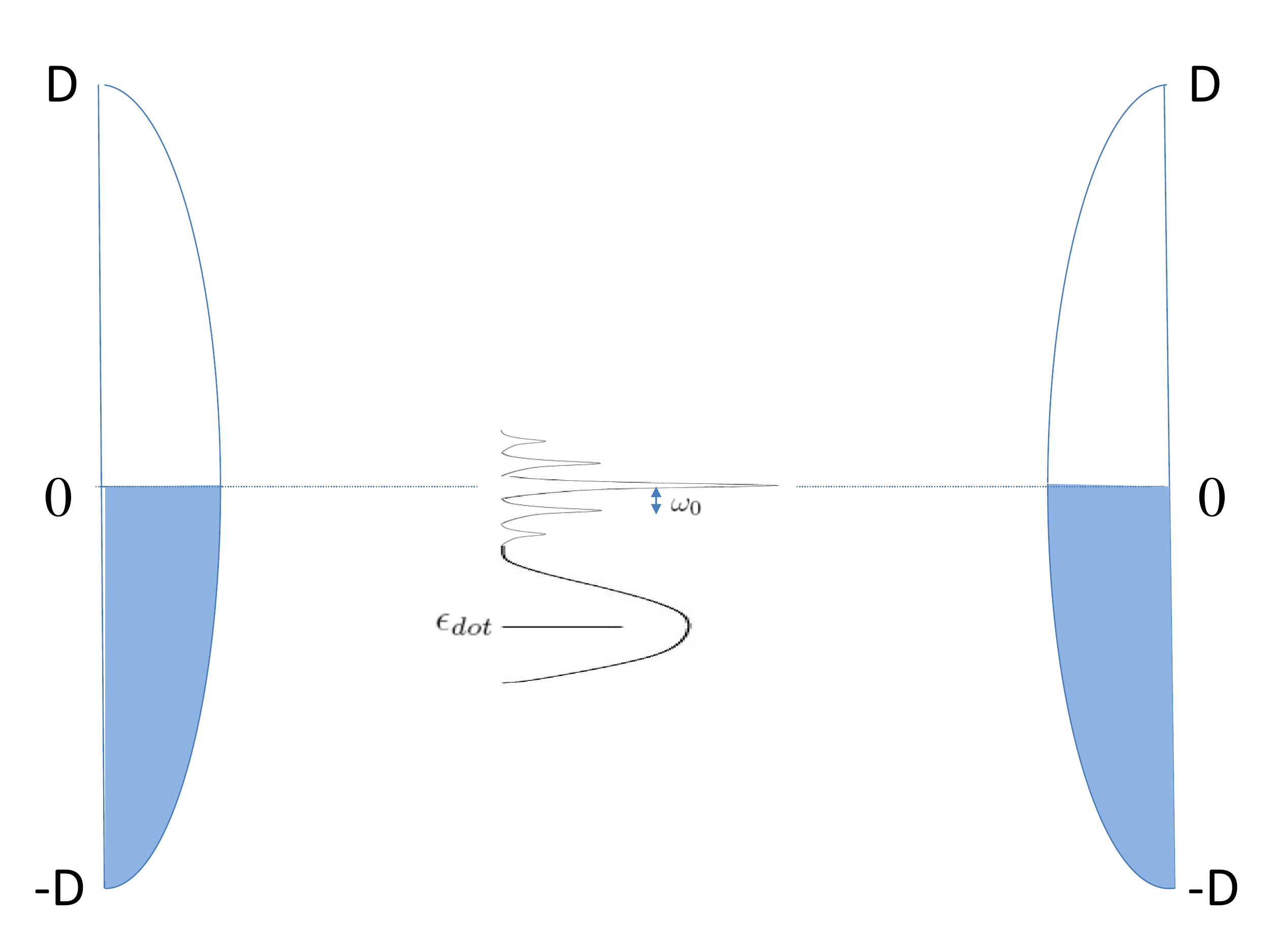}}
\caption{
This figure depicts the density of states of the left and right 
contacts alongside with the density of states of the quantum dot
in the final state schematically.
}
\label{Schem}
\end{figure}

In this paper, we will investigate a scenario 
in which a molecular quantum dot is suddenly shifted
from a position well below the Fermi level of the
leads to a position where the Kondo effect is present by
means of a gate voltage. Previous studies unambiguously
demonstated that the strong electron-electron interactions 
give rise to a Kondo resonance pinned to the Fermi level
of the contacts decorated with sidebands on each side
due to the electron-phonon coupling. However, little 
is known about the real time electron dynamics of this
system as a response to abrupt perturbations except a
recent study using the mean field approximation for a 
spinless model \cite{RiwaretAl09PRB}. The set-up of the 
system under consideration is shown schematically in 
Fig.~\ref{Schem}. The goal of this paper is to fill 
this gap by reporting the instantaneous conductance 
after an interacting vibronic dot's energy level has 
been moved to its final position.

\section{Theory}

We model this device by a single spin degenerate level
of energy $\epsilon_{dot}$ attached to leads through 
tunnel barriers and coupled to a single phonon mode. 
This model is referred as Holstein Hamiltonian. In this 
paper, we will be concerned with the strong electron-phonon
coupling regime where the electron-phonon term can be 
discarded with a canonical transformation. We carry out 
the auxiliary boson transformation for the resulting 
Hamiltonian where the ordinary electron operator on the 
dot is rewritten in terms of a massless boson operator 
and a pseudofermion operator. $U$ $\rightarrow \infty$ 
limit is obtained by imposing the condition that the 
sum of the number of bosons and the pseudofermions is 
equal to unity.

The aforementioned Hamiltonian has three pieces describing 
the contacts, quantum dot and the tunneling process 
between them and it can be written as 
\begin{equation}
H(t)=H_C+H_D(t)+H_T(t)
\end{equation}
where
\begin{eqnarray}
H_C &=& \sum_{k\alpha\sigma}(\epsilon_{k\alpha}-\mu_{\alpha})c^{\dagger}_{k\alpha\sigma}c_{k\alpha\sigma} \nonumber \\
H_D(t)&=& \sum_{\sigma}[\epsilon_{dot}(t)+\lambda(a+a^{\dagger})]d^{\dagger}_{\sigma}d_{\sigma}
+Ud^{\dagger}_{\uparrow}d_{\uparrow}d^{\dagger}_{\downarrow}d_{\downarrow} \nonumber \\
& & +\omega_0 a^{\dagger}a \nonumber \\
H_T(t)&=& \sum_{k\alpha\sigma}(V_{k\alpha}(t)c^{\dagger}_{k\alpha\sigma}d_{\sigma}+h.c.).
\end{eqnarray}

In this Hamiltonian, $d_{\sigma}^{\dagger}(d_{\sigma})$ 
and $c_{k\alpha\sigma}^{\dagger}(c_{k\alpha\sigma})$ with 
$\alpha$=L,R create(annihilate) an electron of spin $\sigma$ 
in the quantum dot and in the left(L) and right(R) leads 
respectively. $V_{k\alpha}$ and $\mu_{\alpha}$ are the tunneling 
amplitudes and chemical potentials for the left and the right 
leads. $a^{\dag}(a)$ creates(annihilates) a phonon within the 
quantum dot. $\lambda$ is the strength of the electron-phonon 
coupling and $\omega_0$ is the phonon frequency. In this paper, 
we will use atomic units with $\hbar=k_B=e=1$.

We will take the hopping matrix elements to be equal 
with no explicit time and energy dependence. Under 
this assumption, broadening of the dot level can be 
parameterized as $\Gamma(\epsilon)=\bar{\Gamma}\rho(\epsilon)$ 
where $\bar{\Gamma}$ is a constant given by 
$\bar{\Gamma}=2\pi|V(\epsilon_f)|^2$ and
$\rho(\epsilon)$ is the density of states function 
of the contacts. We will use parabolic density of 
states in both contacts with same bandwidth.

If the electron-phonon coupling is weak, electron-phonon 
coupling term can be treated perturbatively. In this paper, 
we will be concerned with the opposite case, where 
electron-phonon coupling is sufficiently strong compared 
to the tunnel couplings. In this regime, perturbative 
solution fails and a suitable canonical transformation 
must be applied to eliminate the electron-phonon coupling 
term. The most versatile choice is the unitary Lang-Firsov 
canonical transformation \cite{LangetAl63JETP} given by
\begin{equation}
S=exp \left[\frac{\lambda}{\omega_0}\sum_{\alpha}d^{\dagger}_{\sigma}d_{\sigma}(a^{\dagger}-a)\right].
\end{equation}
This transformation gives 
\begin{eqnarray}
SaS^{\dagger}&=& a-\frac{\lambda}{\omega_0}\sum_{\sigma}d^{\dagger}_{\sigma}d_{\sigma} \nonumber \\
S d_{\sigma}S^{\dagger} &=& d_{\sigma} X
\end{eqnarray}
where the operator $X$ is given by
\begin{equation}
X=exp \left[-\frac{\lambda}{\omega_0}(a^{\dagger}-a)\right].
\end{equation}
The electron operators $d_{\sigma}(d^{\dagger}_{\sigma})$ in 
$H_{T}(t)$ are multiplied with $X(X^{\dagger})$ as a result of 
this transformation indicating that the tunneling electrons 
create and destroy a phonon cloud. This leads to polaron 
formation at the junction.

Under this canonical transformation, dot Hamiltonian turns into
\begin{eqnarray}
\overline{H}_D(t) &=& S H_D(t) S^{\dagger} \nonumber \\
&=&\sum_{\sigma}\left(\epsilon_{dot}(t)-\frac{\lambda^2}{\omega_0} \right) d^{\dagger}_{\sigma}d_{\sigma} \nonumber \\
& &+\left(U-\frac{2\lambda^2}{\omega_0}\right)d^{\dagger}_{\uparrow}d_{\uparrow}d^{\dagger}_{\downarrow}d_{\downarrow}
+\omega_0 a^{\dagger}a.
\end{eqnarray}
It is clear from this result that the dot level and the 
Hubbard interaction strengths are renormalized as 
$\overline{\epsilon}_{dot}(t)=\epsilon_{dot}(t)-(\lambda^2/\omega_0)$
and $\overline{U}=U-(2\lambda^2/\omega_0)$ respectively 
as a result of this transformation.

In the strong electron-phonon coupling regime, i.e. when a 
polaron is formed at the junction, mean-field theory can 
be applied and the expectation value of operator X given by
\begin{equation}
<X>=exp \left[-\frac{\lambda^2}{\omega_0^2}\left(N_{ph.}+\frac{1}{2}\right)\right]
\end{equation}
can serve as a substitute for the operator.

Even though $U$ has been renormalized to $\overline{U}=U-(2\lambda^2/\omega_0)$, 
it is still positive and it overwhelms the line width $\Gamma$ for 
realistic systems. Hence, typically one takes $\overline{U} \rightarrow \infty$. 
This forbids double occupancy of the dot level. The downside of this 
advantage is that the standard diagrammatic techniques are not applicable 
anymore. This problem can be circumvented by introducing a massless boson 
operator and pseudofermion operator on the molecule. The original 
electron operators can be written in terms of these operators as 
\begin{eqnarray}
d_{\sigma}(t)&=& b^{\dagger}(t)f_{\sigma}(t) \nonumber \\
d^{\dagger}_{\sigma}(t) &=& f^{\dagger}_{\sigma}(t)b(t)
\end{eqnarray}
subject to the requirement
\begin{equation}
Q=b^{\dagger}b+\sum_{\sigma}f^{\dagger}_{\sigma}f_{\sigma}=1,
\end{equation}
which ensures the single occupancy of the dot level.
The resulting slave boson Hamiltonian becomes  
\begin{eqnarray}
\overline{H}(t)&=& \sum_{k\alpha\sigma}(\epsilon_{k\alpha}-\mu_{\alpha})c^{\dagger}_{k\alpha\sigma}c_{k\alpha\sigma}
+ \sum_{\sigma} \overline{\epsilon}_{dot}f^{\dagger}_{\sigma}f_{\sigma}+\omega_0 a^{\dagger}a \nonumber \\
& & +\sum_{k\alpha\sigma}(\tilde{V}_{k\alpha}(t)c^{\dagger}_{k\alpha\sigma}f_{\sigma}b^{\dagger}+h.c.),
\end{eqnarray}
where the tunnel coupling and dot level are renormalized as
$\tilde{V}_{k\alpha}(t)=V_{k\alpha}(t)<X>$ and 
$\overline{\epsilon}_{dot}(t)=\epsilon_{dot}(t)-(\lambda^2/\omega_0)$
respectively.

We then invoke the well established non-crossing approximation(NCA) 
to determine the pseudofermion and slave boson self-energies. NCA 
has been shown to give accurate results for dynamical quantities save 
for temperatures below $T/T_K\approx$0.1 or finite magnetic fields. 
These two problematic cases will not be dwelled on in this paper.

The net current flowing through the device can be calculated          
from the resulting Green's functions. We will write  the net current 
as $I(t)=I_L(t)-I_R(t)$, where $I_L(t)(I_R(t))$ represents the 
net current from the left(right) contact through the left(right) 
barrier to the quantum dot. The most general expression for the 
net current \cite{JauhoetAl94PRB} has previously been derived 
by using pseudofermion and slave boson Green's functions \cite{GokeretAl07JPCM}. 
Here, we will adapt it to the present situation. For symmetric coupling, 
the net current is reduced to
\begin{equation}
I(t)=-2\bar{\Gamma} Im(\int_{-\infty}^{t} dt_1 G^R(t,t_1)(f_{L}(t-t_1)-f_{R}(t-t_1))) 
\end{equation} 
and the retarded Green function can be expressed as
\begin{eqnarray}
G^R(t,t_1)&=&-i\theta(t-t_1)[<d_{\sigma}(t)d^{\dagger}_{\sigma}(t_1)><X(t)X^{\dagger}(t_1)> \nonumber \\
& &+<d^{\dagger}_{\sigma}(t_1)d_{\sigma}(t)><X^{\dagger}(t_1)X(t)>].
\end{eqnarray}

It turns out that the correlators of operators $X(t)$ and 
$X^{\dagger}(t_1)$ can be evaluated precisely. They are given 
by $<X(t)X^{\dagger}(t_1)>=e^{\phi(t_1-t)}$ and 
$<X^{\dagger}(t_1)X(t)>=e^{\phi(t-t_1)}$. In these 
expressions, the phase factor is given by 
\begin{eqnarray}
\phi(t_1-t)&=& -g[N_{ph.}(1-e^{-i\omega_0 (t_1-t)})+ \nonumber \\
& &(N_{ph.}+1)(1-e^{i\omega_0(t_1-t)})],
\end{eqnarray}
where $g$ is defined as  $g=\frac{\lambda^2}{\omega_0^2}$ 
and $N_{ph.}$, given by Bose-Einstein distribution 
$N_{ph.}=\frac{1}{e^{\frac{\hbar\omega_0}{k_B T}}-1}$ 
function, represents the average number of phonons for 
temperature $T$ and phonon frequency $\omega_0$. It has 
been pointed out before that the correlators are approximately 
equal only at sufficiently high temperatures such that $N_{ph.}$ 
is much larger than unity. At low temperatures, taking them 
equal may produce erroneous results such as producing phonon 
sidebands only on one side of the main Kondo peak. Therefore, 
in this paper we will not resort to such an approximation 
and keep the correlators in separate terms.

The retarded Green's function can then be written as 
\begin{eqnarray}
G^R(t,t_1)&=& -i\theta(t-t_1)[<d_{\sigma}(t)d^{\dagger}_{\sigma}(t_1)>e^{\phi(t_1-t)} \nonumber\\
& & +<d^{\dagger}_{\sigma}(t_1)d_{\sigma}(t)>e^{\phi(t-t_1)}] \nonumber \\
&=& -i\theta(t-t_1)[G^>_{real}(t,t_1)e^{\phi(t_1-t)} \nonumber\\
& & +G^<_{real}(t,t_1)e^{\phi(t-t_1)}].
\end{eqnarray}
Using the slave boson and pseudofermion decomposition 
of the original fermion operators on the molecule, 
the retarded Green's function can be recasted as
\begin{eqnarray}
G^R(t,t_1) &=& -i\theta(t-t_1)[G^R_{pseudo}(t,t_1)B^<(t_1,t)e^{\phi(t_1-t)} \nonumber \\
& & +G^<_{pseudo}(t,t_1)B^{R}(t_1,t)e^{\phi(t-t_1)}]
\end{eqnarray}

The real-time coupled integro-differential Dyson 
equations for the retarded and less than Green's 
functions are computed in a cartesian two-dimensional grid.
The values are stored in a matrix and the matrix is 
propagated diagonally in time. The phase factors remain 
attached to the pseudofermion Green's functions during this
procedure in order to incorporate the phonon effects 
properly \cite{WerneretAl07PRL}. The instantaneous 
conductance is obtained by performing an integration 
over the lowest row of the matrix using the expression
\begin{eqnarray}
I(t)&=&2\bar{\Gamma} Re ( \int_{-\infty}^{t} dt_1
(G_{pseudo}^{R}(t,t_1)B^{<}(t_1,t)e^{\phi(t_1-t)} \nonumber \\
& &+G_{pseudo}^{<}(t,t_1)B^{R}(t_1,t)e^{\phi(t-t_1)})\times \nonumber \\
& &(f_{L}(t-t_1)-f_{R}(t-t_1))),
\end{eqnarray}
where $f_L(t-t_1)$ and $f_R(t-t_1)$ are the convolution 
of the density of states function with the Fermi-Dirac 
distribution \cite{GokeretAl07JPCM}. The conductance 
$G$ is equal to the current divided by the bias voltage.
A comprehensive description of our numerical implementation 
has been published previously \cite{ShaoetAl194PRB,IzmaylovetAl06JPCM}.

An exquisite many-body state called the Kondo effect 
arises when the dot level is positioned below the Fermi 
energy at sufficiently low temperatures. The net spin 
localized within the dot and the Fermi sea of electrons 
in the contacts hybridize to form a spin singlet. This 
results in a sharp resonance fixed to the Fermi levels 
of the contacts in the dot density of states. The 
linewidth of the Kondo resonance can be approximated by 
an energy scale $T_K$ (Kondo temperature) given by
\begin{equation}
T_K \propto \left(\frac{D\Gamma}{4}\right)^\frac{1}{2}
\exp\left(-\frac{\pi|\epsilon_{\rm dot}|}{\Gamma}\right),
\label{tkondo}
\end{equation}
where $D$ is a high energy cutoff equal to half bandwidth
of the conduction electrons and $\Gamma$ corresponds to
the value of coupling between the dot and the contacts 
$\Gamma(\epsilon)$ at $\epsilon=\epsilon_F$.

Our aim in this paper is to theoretically investigate 
a case in which the dot level is displaced from its 
equilibrium level abruptly by means of a gate voltage. 
We will be particularly interested in a system which 
has been studied before in the absence of any 
electron-phonon coupling. Its dot level is abruptly 
moved from $\epsilon_{dot}=-5\Gamma$ to 
$\epsilon_{dot}=-2\Gamma$ at $t=0$ where 
$\Gamma=\bar{\Gamma}\rho(\epsilon_f)$. We will 
report the instantaneous conductance right 
after the dot is moved to its final position.
In the following discussion, we will take 
the phonon frequency and the bandwidth 
as $\omega_0=0.06\Gamma$ and $D=9\Gamma$
respectively with $\Gamma$=0.4 eV.

\section{Results}

\begin{figure}[htb]
\centerline{\includegraphics[angle=0,width=8.0cm,height=6cm]{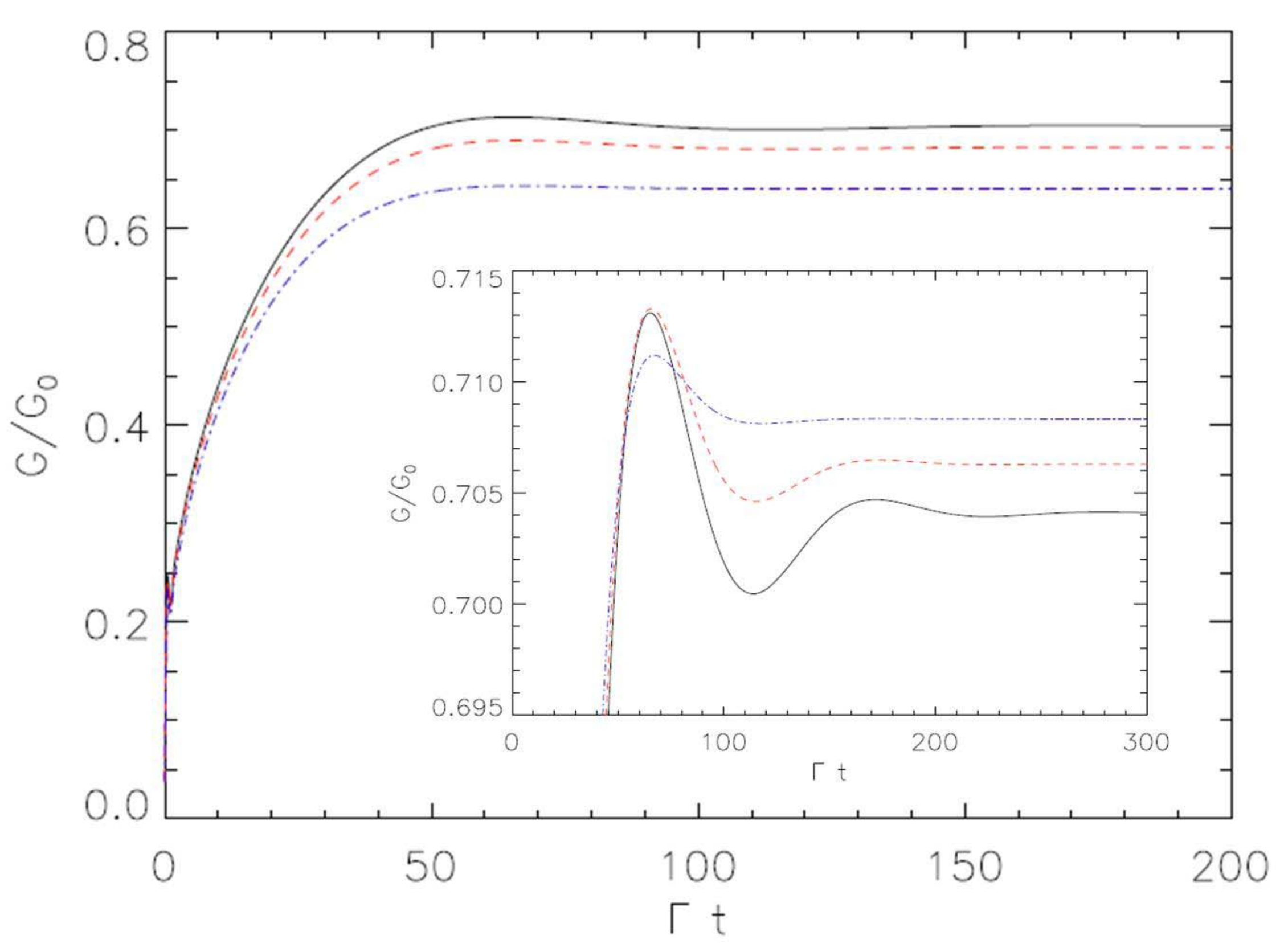}}
\caption{
Black(solid), red(dashed) and blue(dash dotted)
curves show instantaneous conductance results as a
function of time for T=0.0009$\Gamma$, T=0.0010$\Gamma$
and $T$=0.0012$\Gamma$ respectively with $g$=2.25 after
the dot level has been switched to its final position.
The inset is the magnified version of the main panel.
The curves for high temperatures in the inset have been 
shifted to enable camparison of the amplitudes.
}
\label{Fig2}
\end{figure}

We begin our analysis with the instantaneous
conductance results immediately after the dot
level has been switched to its final position.
The results shown in Fig.~\ref{Fig2} correspond
to three different temperatures for a constant
nonzero electron-phonon coupling $g$. Instantaneous
conductance results in the absence of any 
electron-phonon coupling has been reported 
previously \cite{PlihaletAl05PRB}. 
In Fig.~\ref{Fig2}, short timescale corresponds to 
$\Gamma$ t $<$ 10. The conductance oscillations
related to the charge transfer in this timescale
have already been analyzed in detail 
\cite{PlihaletAl05PRB,MuhlbacheretAl08PRL},
therefore we will not elaborate on them here.
The Kondo timescale occurs between the end 
of the short timescale and attainment of a 
plateau by the instantaneous conductance. Kondo 
resonance starts developing in this timescale. It
takes place roughly between 10 $< \Gamma$ t $<$ 60
in Fig.~\ref{Fig2}.

The first key feature that should be mentioned
regarding Fig.~\ref{Fig2} is that the steady
state conductances (i.e. long time limit) are
smaller than the case without any electron-phonon
coupling for all temperatures as expected \cite{YangetAl10PLA}. 
This anticipation stems from the fact that the 
electron-phonon coupling gradually quenches the 
Kondo effect \cite{GalperinetAl07PRB}. This suppression 
has been explained with dephasing due to electron-phonon 
coupling and downward shift of the energy level as
a result of phonon reorganization. The second effect
is an obvious consequence of the renormalization
of the dot level.

The second and more subtle effect is the sinusoidal
oscillation of the current in the Kondo timescale.
This effect is difficult to see in the main panel of 
Fig.~\ref{Fig2}, therefore in the inset, we show the 
magnification of the main panel with shifted conductance 
curves such that they overlap at the onset of oscillations. 
In the inset of Fig.~\ref{Fig2}, it is clear that the
oscillation frequency is the same for all temperatures.
Moreover, the amplitude of the oscillations decreases
as the ambient temperature increases. It turns out
that the oscillation frequency is equal to the 
phonon frequency $\omega_0$.

\begin{figure}[htb]
\centerline{\includegraphics[angle=90,width=8.5cm,height=6.4cm]{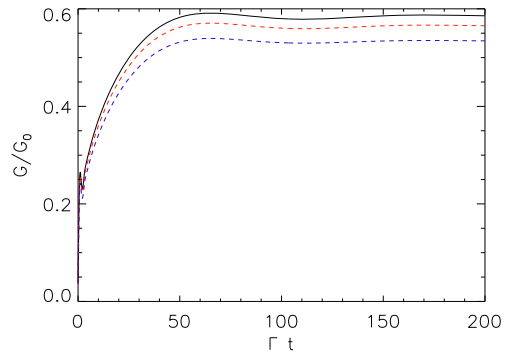}}
\caption{
Black(solid), red(dashed) and blue(dash dotted)
curves show instantaneous conductance results as a
function of time for T=0.0009$\Gamma$, T=0.0010$\Gamma$
and $T$=0.0012$\Gamma$ respectively with $g$=3.75 after
the dot level has been switched to its final position.
}
\label{Fig3}
\end{figure}

In order to test the electron-phonon coupling
strength dependency of the oscillation frequency, 
we increased $g$ and performed the previous 
calculation at the same temperatures. The result 
is shown in Fig.~\ref{Fig3}. First, the steady state 
conductance is lower for all temperatures for 
larger $g$. This is expected due to dephasing and 
downward level shift as pointed out above. On 
the other hand, conductance oscillations once more
take place with a frequency equal to the phonon 
frequency $\omega_0$. We should mention that we 
increased $g$ in Fig.~\ref{Fig3} by increasing 
$\lambda$ and keeping $\omega_0$ constant with 
respect to Fig.~\ref{Fig2} to facilitate a direct 
comparison. It must be noted that changing the value 
of $\omega_0$ in our calculations again yields the 
oscillation frequency of the conductance as $\omega_0$.
We checked that the oscillation frequency remains 
pinned to $\omega_0$ for all $g$ values.

\begin{figure}[htb]
\centerline{\includegraphics[angle=90,width=8.5cm,height=6.4cm]{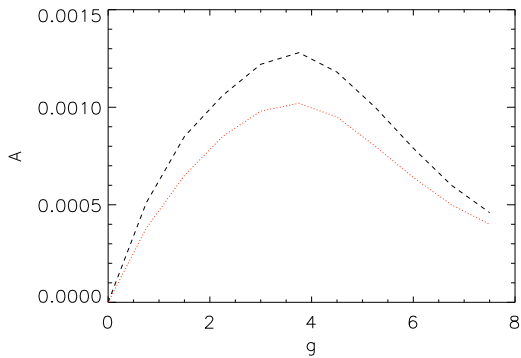}}
\caption{
Black(dashed) and red(dotted) curves show amplitude
of the second peak in conductance oscillation as a function
of the electron-phonon coupling strength $g$ for     
$T$=0.0009$\Gamma$ and $T$=0.0010$\Gamma$ respectively.
The amplitudes have been measured with respect to the
steady state (i.e. $\Gamma$ t $\rightarrow \infty$)            
conductance values.  
}
\label{Fig4}
\end{figure}

We now want to study the amplitude of oscillations 
for various electron-phonon coupling strengths and 
temperatures in a systematic fashion. Fig.~\ref{Fig4} 
shows the behaviour of amplitudes as a function
of electron-phonon coupling strength at two 
different temperatures. The amplitude is zero 
at $g$=0 regardless of the temperature but 
it starts to increase gradually until it reaches
a maximum before $g$=4. It then starts decreasing 
and it again approaches zero for large $g$ values.
Meanwhile, the amplitude of oscillation is always
larger at lower temperature for a given $g$. We
verified that this conclusion holds for other
temperatures as well.

It is of great interest to be able to provide a 
microscopic description for this peculiar 
behaviour of the transient current. One needs 
to resort to the spectral function of the dot 
to this end. This can be obtained by taking the 
Fourier transform of the retarded Green's 
function. It has been shown before that  
the phonon sidebands are separated from 
the main Kondo peak by an integer multiple
of phonon frequency $\omega_0$ 
\cite{ChenetAl05PRB,GalperinetAl06PRB,WangetAl07PRB,PaaskeetAl05PRL}. 
We propose that the sinusoidal oscillations seen 
in the transient current in the Kondo timescale
are a result of an interference between the
main Kondo peak and its phonon sidebands.
For this reason, the frequency of oscillation
was found to be equal to $\omega_0$. 
For a given $g$ parameter, the amplitude 
of oscillations increases as the temperature 
decreases because both the main Kondo peak 
and its phonon satellites are more developed 
at lower temperatures leading to stronger 
interference.

On the other hand, sweeping electron-phonon 
coupling strength at constant T gives rise to a 
more complicated situation as seen in Fig.~\ref{Fig4}. 
Even though the main Kondo peak is developed most 
fully for $g$=0 at a given temperature $T$ 
resulting in largest steady state conductance, 
no oscillation was detected in infinitesimal 
bias in Kondo timescale \cite{PlihaletAl05PRB}
because the phonon sidebands are completely
absent in this case. Consequently, interference
is nonexistent and oscillation amplitude
is zero. For small $g$ values, amplitude of the 
oscillations gradually increases with $g$ 
because while the main Kondo peak pinned to the 
Fermi level gets inhibited a bit, its phonon 
sidebands become slightly more pronounced  
\cite{YangetAl10EPL}. This naturally leads to 
stronger interference. Note that for $g<$0.5,
the system is no longer in strong coupling 
regime. Still, our simulations show a consistent
behaviour in this limit. The behaviour 
changes for moderate values of $g$. The 
amplitudes reach a maximum around $g$=4
and then start decreasing for larger $g$
values since destruction sets in for all peaks.  
Unsurprisingly, the amplitude goes to zero 
in large $g$ limit where the Kondo effect 
disappears completely and none of the peaks 
survives.    

\begin{figure}[htb]
\centerline{\includegraphics[angle=0,width=7.4cm,height=5.6cm]{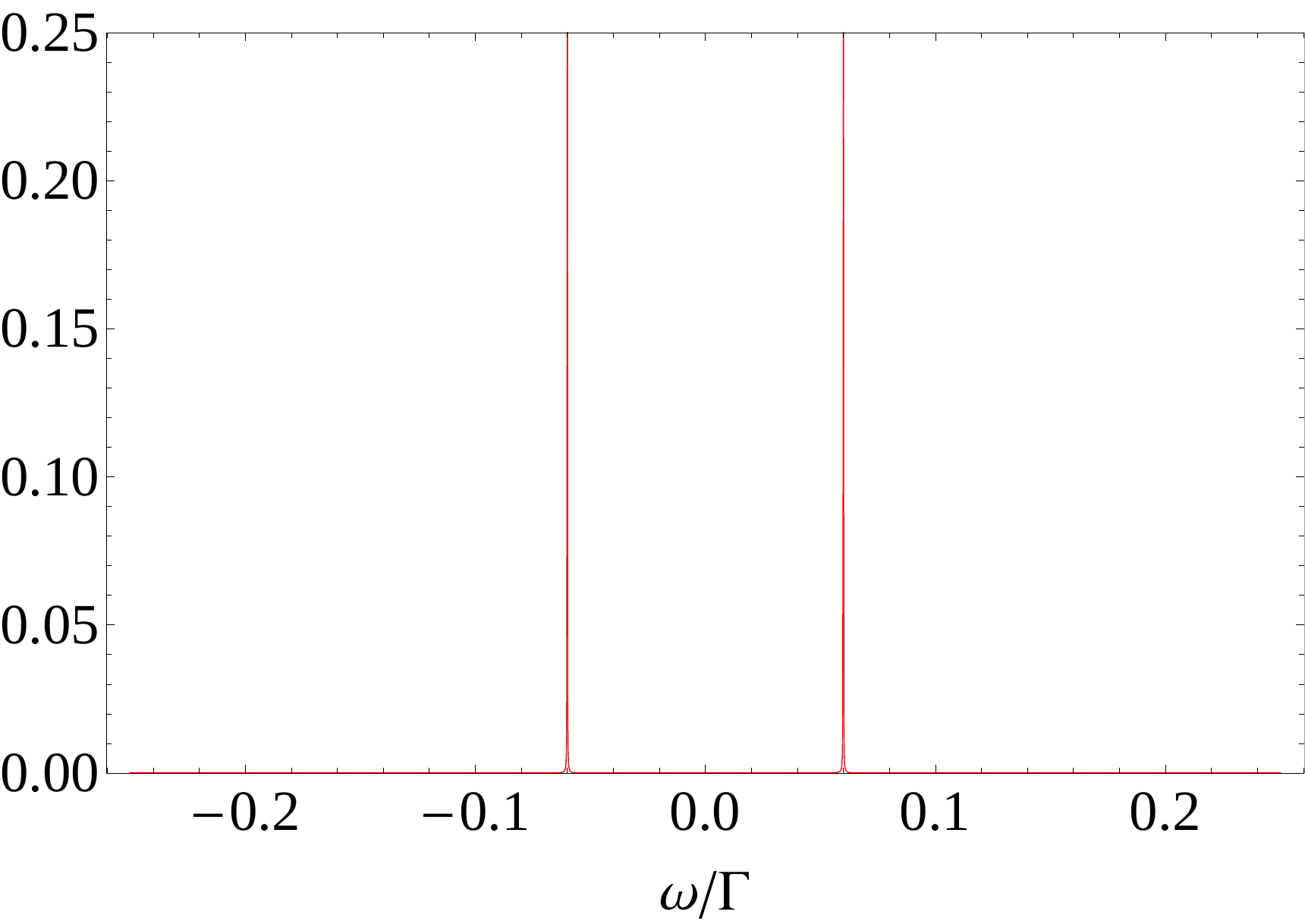}}
\caption{
This figure shows the Fourier transform of the
instantaneous conductance in the long timescale
in infinitesimal bias. 
}
\label{Fig5}
\end{figure}

It is important to note that the oscillations 
take place exclusively with $\omega_0$ in
infinitesimal bias and all the integer 
multiples of this frequency are completely 
absent as one can easily see from the Fourier 
transform of the instantaneous conductance 
in the long timescale shown in Fig.~\ref{Fig5}. 
This result is somewhat unexpected since the 
main Kondo peak should be interfering with 
other sidebands as well. This issue can be
explained easily from an intuitive point of view. 
As one can see in Fig.~\ref{Schem} schematically 
and confirm by taking the Fourier transform of 
the retarded Green's function, sidebands start 
getting drastically smaller away from the main 
Kondo peak resulting in less interference with 
it. Therefore, the oscillation amplitudes 
associated with these peaks is negligible 
compared to the one associated with the first 
sideband leading to the absence of other frequencies.

Before we conclude, we would like to address 
the effect of finite bias on the above reported 
time dependent conductance. First, it is well 
known that the bias would split the main elastic 
Kondo peak into two each of which are pinned to 
the Fermi level of the contacts. These split 
Kondo resonances would induce SKP oscillations 
in transient current with a frequency equal to 
the bias $V$ in the absence of any electron-phonon 
coupling \cite{PlihaletAl05PRB}. Therefore, the 
effect of bias on the above system is two fold. 
First, steady state conductance would be lower 
than the infinitesimal bias case since the 
Kondo effect is gradually destroyed with bias. 
In fact, steady state investigation of a system 
with both strong electron-electron and electron-phonon 
interaction in finite bias showed that the 
differential conductance exhibits enhancements 
when the bias is an integer multiple of the 
phonon frequency $\omega_0$ \cite{PaaskeetAl05PRL}. 
This is due to the fact that the inelastic phonon 
sidebands overlap with the split Kondo peaks. 
Indeed, this behaviour is reminiscent of the time 
averaged conductance of an ac driven quantum dot 
\cite{Goker08SSC}.

\begin{figure}[htb]
\centerline{\includegraphics[angle=0,width=7.8cm,height=5.6cm]{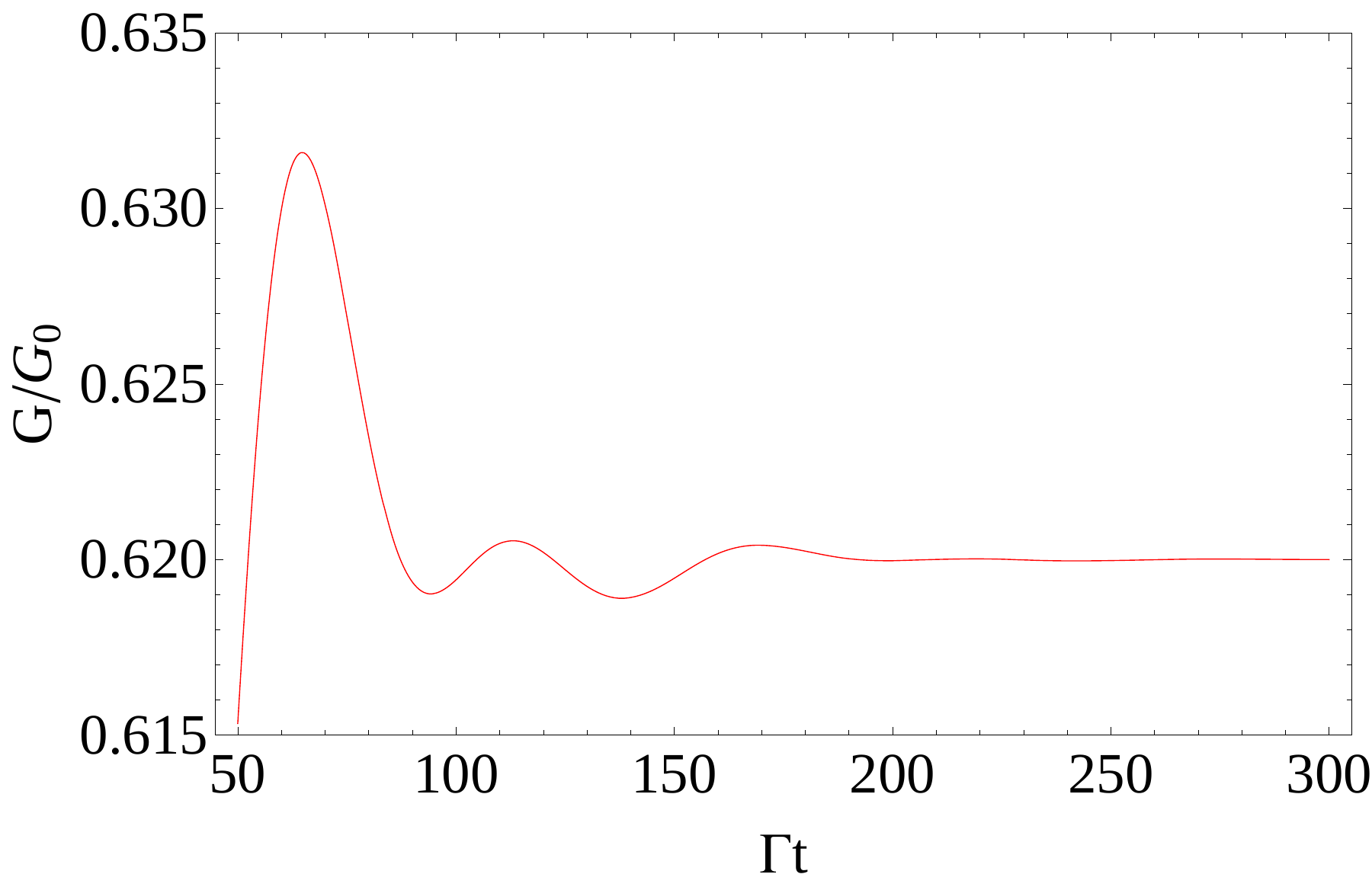}}
\caption{
This figure shows the instantaneous conductance as a
function of time for T=0.0009$\Gamma$ with $g$=2.25 
and voltage bias of V=0.03$\Gamma$ after the dot 
level has been switched to its final position.
}
\label{Fig6}
\end{figure}

In time dependent case, there are two interference 
effects taking place simultaneously, namely between 
the split main Kondo peaks and between each split 
main Kondo peak and its phonon sidebands. Obviously, 
this results in two distinct oscillation frequencies. 
This beating gives rise to more complicated oscillations 
in the transient current as seen in Fig.~\ref{Fig6}. 
When we take the Fourier transform of the transient 
current, we are able to identify these two distinct 
frequencies corresponding to the bias $V$ and the 
phonon frequency $\omega_0$. This result unambiguously 
confirms the validity of above interpretation. When the 
bias is an integer multiple of $\omega_0$, behaviour of 
the transient current is quite similar to zero bias case. 
In this case, only one frequency survives and it is equal 
to $\omega_0$.

\section{Conclusions}

In conclusion, the non-crossing approximation was 
utilized in this paper to investigate the combined 
effect of strong electron-electron and electron-phonon 
interaction on instantaneous conductance in single 
molecule transistor when the dot level is suddenly 
shifted to a position where the Kondo resonance is 
present. Our results clearly showed that the instantaneous 
conductance displays decaying sinusoidal oscillations 
in the long timescale in infinitesimal bias. This fact
set these novel oscillations apart from the previously
predicted split Kondo peak oscillations \cite{PlihaletAl05PRB}.

Investigation of the amplitude of these oscillations 
indicated that it sensitively depends on the ambient 
temperature and the electron-phonon coupling strength. 
Moreover, the frequency of these oscillations was found 
to be precisely equal to the phonon frequency upon
taking the Fourier transform of the instantaneous
conductance in the long timescale. Based on 
these observations and investigation of the density of 
states of the quantum dot, we proposed that the origin 
of this novel phenomenon can be attributed to the 
interference between the main elastic Kondo peak and 
its phonon sidebands. We also uncovered that finite 
source-drain bias results in more complicated
fluctuation patterns in the instantaneous conductance.
Two distinct frequencies corresponding to the phonon
frequency $\omega_0$ and bias voltage $V$ 
play role in this case.

We believe the phenomenon discussed in this paper can be 
observed with present day ultrafast transient experimental 
techniques \cite{Teradaetal10JPCM} since it takes place in 
the longest timescale which typically occurs on the 
order of tens of picoseconds. Besides, the theory presented 
here depicts a fairly realistic assessment of the single 
molecule junctions as it takes into account both Coulomb 
and vibrational interactions that are ubiquitous for real 
molecules, thus we hope to invigorate this field by 
motivating a new experiment with the predictions in this paper.

\bibliographystyle{iopams} 
\bibliography{gen}

\end{document}